\documentclass[final]{svjour2}
\usepackage{graphicx}
\usepackage{rotating}
\usepackage{amssymb}
\usepackage{mathptmx}
\usepackage[numbers]{natbib}
\makeatletter
\journalname{Journal of Low Temperature Physics}

\bibpunct{}{}{,}{s}{}{,}

\begin{document}

\newcommand{\hdblarrow}{H\makebox[0.9ex][l]{$\downdownarrows$}-}
\title{Characterization and physical explanation of energetic particles on Planck HFI instrument}

\author{A.~Catalano$^{10}$
P.~Ade $^{12}$ \and Y.~Atik$^{6}$ \and A.~Benoit$^{7}$ \and E.~Br\'eele$^{1}$ \and J.J.~Bock$^{2,}$$^{9}$ \and P.~Camus$^{7}$ \and M.~Charra$^{6}$ \and B.P.~Crill$^{9}$ \and N.~Coron$^{6}$ \and A.~Coulais$^{11}$ \and F.-X.~D\'esert$^{8}$ \and L.~Fauvet$^{5}$ \and Y.~Giraud-H\'eraud$^{1}$ \and  O.~Guillaudin$^{10}$ \and W.~Holmes$^{9}$ \and W. C.~Jones$^{4}$ \and J.-M.~Lamarre$^{11}$ \and J.~Mac\'{\i}as-P\'erez$^{10}$ \and M.~Martinez$^{6}$ \and A.~Miniussi$^{6}$ \and A.~Monfardini$^{7}$ \and F.~Pajot$^{6}$ \and G.~Patanchon$^{1}$ \and A.~Pelissier$^{10}$ \and M.~Piat$^{1}$ \and J.-L.~Puget$^{6}$ \and C.~Renault$^{10}$ \and C.~Rosset$^{1}$ \and D.~Santos$^{10}$ \and A.~Sauv\'e$^{3}$ \and L.~Spencer$^{12}$ \and R.~Sudiwala$^{12}$}

\institute{1: Astroparticule et Cosmologie, CNRS (UMR 7164), Universit\'e Denis Diderot Paris 7, B\^atiment Condorcet, 10 rue A. Domon et Leonie Duquet, Paris, France, \\ 
2: California Institute of Technology, 1200 E California Blvd, Pasadena, California, USA, \\ 
3: CNRS, IRAP, 9 Av. colonel Roche, BP 44346, 31028 Toulouse Cedex 4, France, \\ 
4: Department of Physics, Princeton University, Princeton, New Jersey, USA , \\ 
5: European Space Agency, ESTEC, Keplerlaan 1, 2201 AZ Noordwijk, The Netherlands, \\ 
6: Institut d'Astrophysique Spatiale, CNRS (UMR8617) Universit\'e Paris-Sud 11, B\^atiment 121, Orsay, France, \\ 
7: Institut N\'eel, CNRS, Universit\'e Joseph Fourier Grenoble I, 25 rue des Martyrs, Grenoble, France, \\ 
8: IPAG: Institut de Plan\'etologie et d'Astrophysique de Grenoble, Universit\'e Joseph Fourier, Grenoble 1/CNRS-INSU, UMR 5274, 38041 Grenoble, France, \\ 
9: Jet Propulsion Laboratory, California Institute of Technology, 4800 Oak Grove Drive, Pasadena, California, USA, \\ 
10: Laboratoire de Physique Subatomique et de Cosmologie, CNRS/IN2P3, Universit\'e Joseph Fourier Grenoble, Institut National Polytechnique de Grenoble, 53 rue des Martyrs, 38026 Grenoble Cedex, France, {\bfseries \email{catalano@lpsc.in2p3.fr}} \\ 
11: LERMA, CNRS, Observatoire de Paris, 61 avenue de l'Observatoire, Paris, France, \\ 
12: School of Physics and Astronomy, Cardiff University, Queens Buildings, The Parade, Cardiff, CF24 3AA, UK.}


\date{07.10.2013}

\maketitle

\begin{abstract}

The Planck High Frequency Instrument (HFI) has been surveying the sky continuously from the second Lagrangian point (L2) between August 2009 and January 2012. It operates with 52 high impedance bolometers cooled at 100mK in a range of frequency between 100 GHz and 1THz with unprecedented sensivity, but strong coupling with cosmic radiation.
At L2, the particle flux is about 5 $cm^{-2} s^{-1}$ and is dominated by protons incident on the spacecraft. Protons with an energy above 40MeV can penetrate the focal plane unit box causing two different effects:  glitches in the raw data from direct interaction of cosmic rays with detectors (producing a data loss of about 15\% at the end of the mission) and thermal drifts in the bolometer plate at 100mK adding non-gaussian noise at frequencies below 0.1Hz. The HFI consortium has made strong efforts in order to correct for this effect on the time ordered data and final Planck maps.
This work intends to give a view of the physical explanation of the glitches observed in the HFI instrument in-flight. To reach this goal, we performed several ground-based experiments using protons and $\alpha$ particles to test the impact of particles on the HFI spare bolometers with a better control of the environmental conditions with respect to the in-flight data.
We have shown that the dominant part of glitches observed in the data comes from the impact of cosmic rays in the silicon die frame supporting the micro-machinced bolometric detectors propagating energy mainly by ballistic phonons and by thermal diffusion. The implications of these results for future satellite missions will be discussed.


\end{abstract}

\section{Introduction}\label{intro}
Planck\cite{planck2013p01} is a project of the European Space Agency (ESA) with instruments provided by two scientific consortia funded by ESA member states (in particular the lead countries France and Italy), with contributions from NASA(USA) and telescope reflectors provided by a collaboration between ESA and a scientific consortium led and funded by Denmark\footnote{http://www.esa.int/Planck}.
It comprises a telescope, two instruments High Frequency Instrument HFI and Low frequency Instrument LFI and a Spacecraft. The High Frequency Instrument (HFI) \cite{Planck2011perf} has be operating with 52 high impedance spiderweb bolometers cooled at 100mK in a range of frequency between 100GHz and 1THz. During the mission the HFI instrument has shown a sensitivity in agreement with requirements but at the same time a strong coupling with cosmic radiation.

Cosmic rays (CRs) \cite{Mewaldt2010}$^,$  \cite{Leske2011} at Lagrangian point L2 are essentially composed by massive particles: about 89\% of protons, 10\% of alpha particles, 1\% are the nuclei of heavier elements and less than 1\% of electrons like beta particles.
The total flux of CRs peaks around 200\,MeV, giving a total proton flux of 3000\,-\,4000\,particles\,m$^{-2}$\,sr$^{-1}$\,s$^{-1}$\,GeV$^{-1}$.
The flux is dominated by galactic CRs which dominates in period of low solar activity. The solar wind decelerates the incoming particles and blocks some of the particles with energies below about 1 GeV. Since the amount of solar wind is not constant due to changes in solar activity, the level of the cosmic ray flux varies with time. This is monitored in Planck satellite by the Standard Radiation Environment Monitor (SREM).
The period following the Planck launch was a period of exceptionally low solar activity, resulting in very weak solar modulation of cosmic rays at 1AU\cite{Mewaldt2010}. The flux of low energy ($200MeVnucleon^{-1}$) nuclei from carbon to iron was four times higher than in the period between 2001 and 2003, and 20\% higher than in previous solar minima over the last 40 years.

In this paper, first we discuss briefly the characteristic of the HFI high impedance bolometers. In section \ref{glitch} we describe the impact of the cosmic ray in the in-flight HFI data. In section \ref{ground} we give the results obtained from ground-based tests which permitted to understand the origins of the different families of the in-flight glitches.  

\begin{figure}[t!]
\begin{center}
\includegraphics[width=10cm, , keepaspectratio]{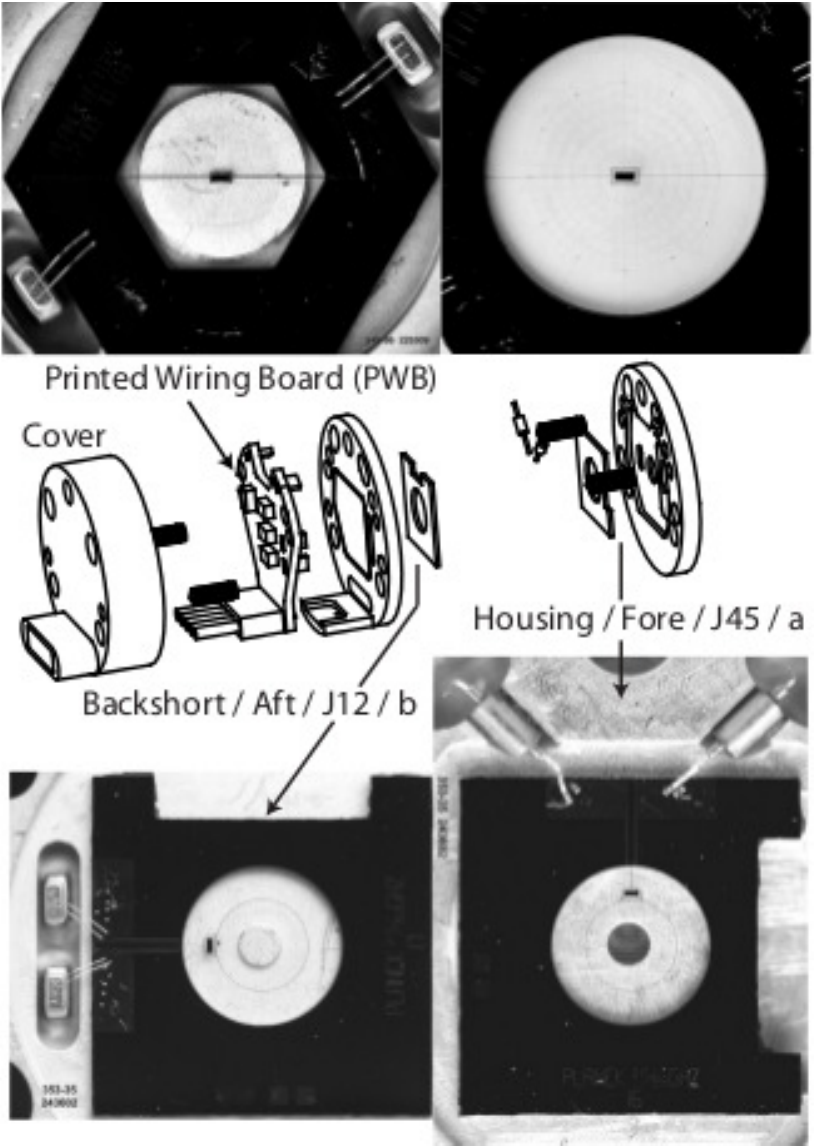}
\end{center}
\caption{Top: SpiderWeb Bolometer (SWB) with the sensitive NTD thermometer on the centre of the spider absorber (zoom of the absorber and the NTD on top right picture.).  Middle: exploded view of a typical HFI bolometer module. Bottom: Polarization Sensitive Bolometer (PSB). Two PSBs are mounted to the same bolometer module. The two pictures show the forward/upper PSBs (PSBb and PSBa).}
\label{fig:bol}
\end{figure}

\section{HFI Bolometers}\label{bolo}

The HFI bolometers are made
of a Neutron Transmutation Doped thermistor 30\,$\mu$m (thickness) $\times$ 100 $\mu$m $\times$ 350\,$\mu$m identical for all these detectors, a
free standing metallized $Si_3N_4$ micromesh supported by $Si_3N_4$ beams (thickness 1 $\mu$m)
 and a silicon die. Other elements that composes a bolometer module have no impact on the goals of this paper so they are not discussed here.
 Details can be found in Holmes et al, 2008\cite{Holmes2008}.
 
 For SWB detectors (spiderweb
bolometer, non-sensitive to polarization), the thermometer
is at the center of the grid while for PSB detectors (polarization sensitive
bolometer) the thermometer is at the edge of the grid
as it can be seen on Fig \ref{fig:bol}. The grid geometry has been
chosen so that it absorbs mm-waves with high efficiency but has a much smaller
physical surface area reducing significantly the cross section to cosmic ray
particles and shorter wavelength photons.

The Silicon die thickness is equal to 350\,$\mu$m, common to
all the bolometers and the surface area is between 0.4 and 0.8\,cm$^{2}$
depending on the working frequency.

\section{HFI glitchology}\label{glitch}

\begin{figure}[t!]
\begin{center}
\includegraphics[width=12cm, , keepaspectratio]{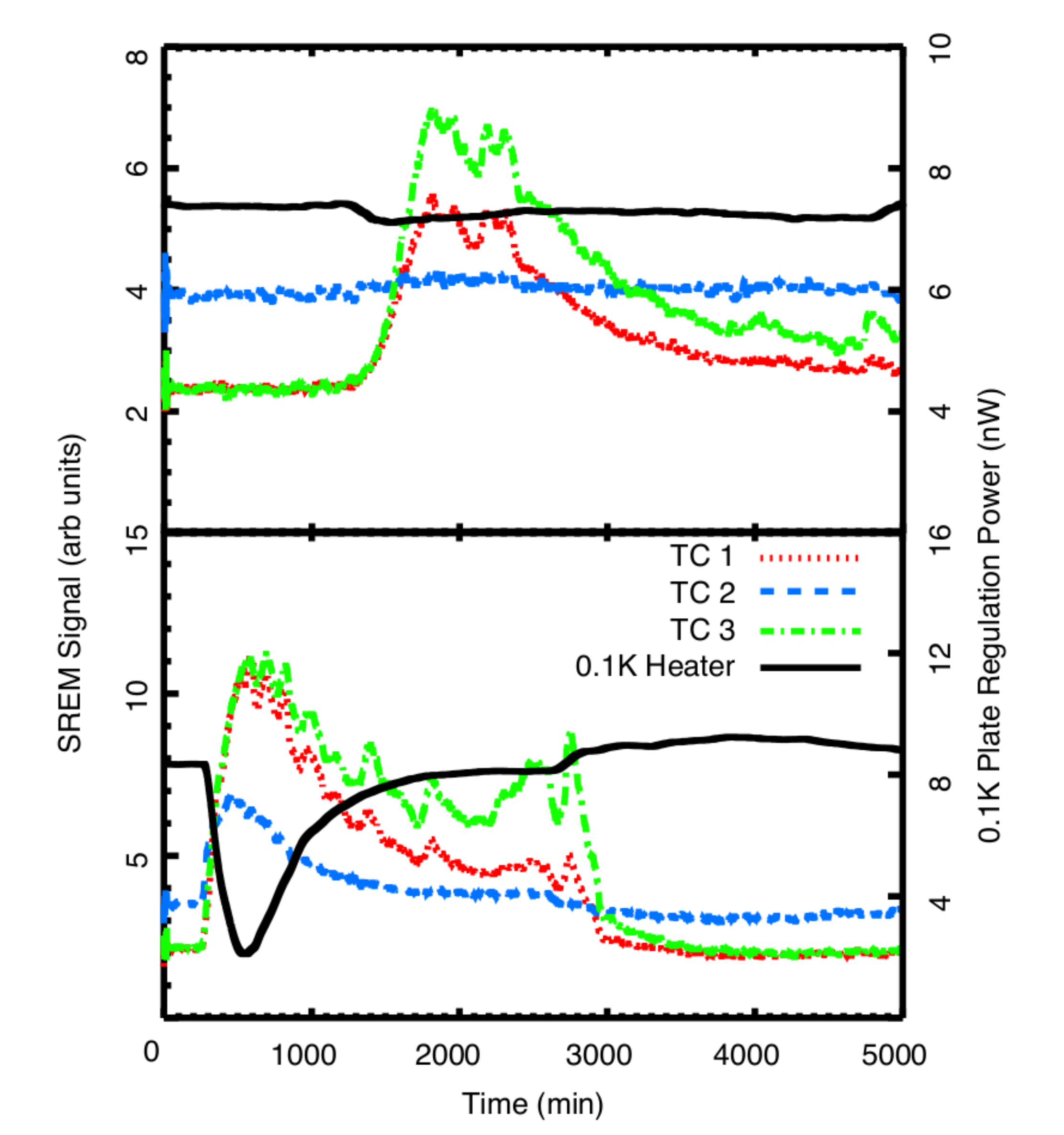}
\end{center}
\caption{Output of SREM diodes D1, D2 and D3 (left axis) and temperature control heater power on the 0.1K plate (right axis) and as a function of time for large solar flares on March 7 (top) and August 4 (bottom) in 2011.}
\label{fig:srem}
\end{figure}

\subsection{The Standard Radiation Environment Monitor}
The flux of CRs is monitored onboard the satellite by the
Standard Radiation Environment Monitor (SREM)\ mounted on the exterior of the Planck spacecraft. SREM consists of three detectors (Diodes D1, D2, D3)  in two detector head configurations \cite{SREM2003}. A total of 15 discriminator levels are available to bin the energy of the detected events. 
Solar flares provided a useful test to correlate the signal measured on the out-side 
of the spacecraft with the SREM to signals due to particle impact on HFI. 

During a solar flare the glitch\footnote{A glitch is the transient effect on the bolometer time ordered data
 associated with a cosmic ray impact somewhere in the detector system.} rate increases and the heater power used to regulate the 0.1K cryogenic stage decreases with the increasing deposited power by the particle flux. In Fig\ \ref{fig:srem} the signal of the three different SREM diodes and the
temperature control heater on 0.1K cryogenic stage for two large solar flares is shown.

 The signals of the heater power and D2 are similar to each other in each flare. However,
the peak signal of each is very different comparing the two flares. In
addition, there is structure in the signal for D1 and D3 that are
not in D2 or the heater power response.  We find that this
correlation holds between the heater power and D2 for all flares.
The diode D2 has the most shielding of the 3 diodes in the SREM,
1.7\,mm of Aluminum and 0.7\,mm of Tantalum which allows only only ions and
protons with energies $>39$\,MeV to pass. The other diodes are shielded by
0.7\,mm of Aluminum for D3 and 1.7\,mm of Aluminum for D1.  This
demonstrates that the spacecraft and instrument surrounding the
bolometers shield particles with energies at least up to 39\,MeV and
all solar electrons.  This is equivalent to the stopping power of
$\sim 1.5$\,cm of Aluminum.

\subsection{Indirect Effect on detectors}

\begin{figure}[b!]
\begin{center}
\includegraphics[width=10cm, , keepaspectratio]{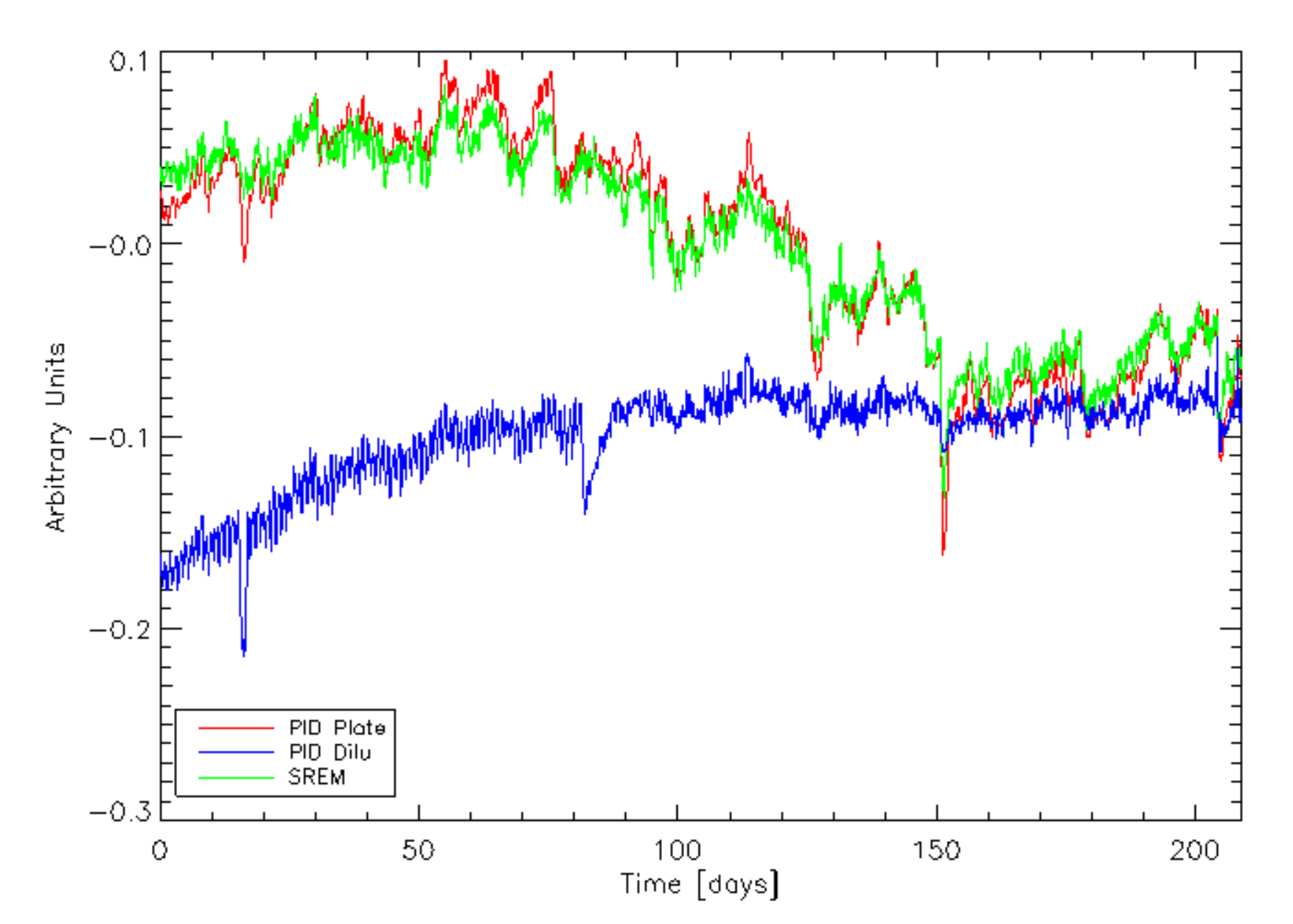}
\includegraphics[width=10cm, keepaspectratio]{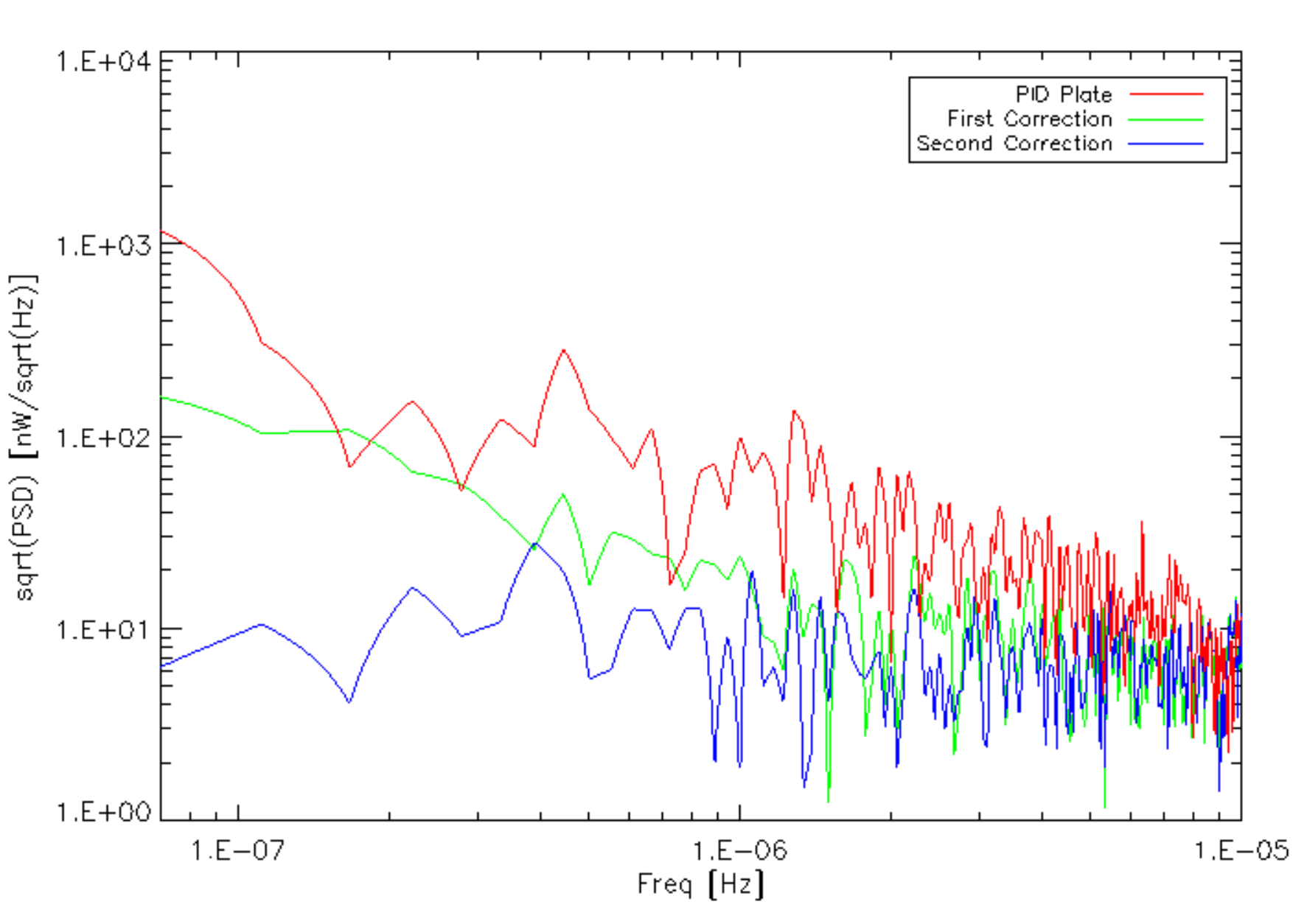}
\end{center}
\caption{Left: correlation between the signal of the SREM (red) and the signal of the active regulation of the temperature of the bolometer plate. Right: corrected data by subtraction of the SREM (first) and dilution fluctuation signal (second).}
\label{fig:ind_eff}
\end{figure}

The very good correlation between the 100mK stage control heater\cite{planck2011ther} and the SREM is shown in Fig \ref{fig:ind_eff} left. The correlation with the control heater of the temperature of
the dilution plate is smaller. Fig \ref{fig:ind_eff} right shows the correction performed by subtracting the data with the SREM (particle contribution) and the fluctuations of the dilution stage (cryogenic contribution). These corrections demonstrate that we have identified all the source of fluctuation in the range of frequencies between 10$^{-7}$ and 10$^{-5}$\,Hz.
In flight, the 100mK temperature fluctuations induced by the
modulation of Galactic cosmic rays dominate, but they do not
affect the signal. In the frequency range between 16 mHz and
300 mHz, excess noise from cosmic rays with respect to ground measurements is
seen on the bolometers and thermometers; however, any common
thermal mode affecting all thermometers and bolometers
that is not fully corrected by the bolometer plate control heaters is 
removed using the dark bolometer signals.
After corrections we find a flat noise spectrum in agreement with the one obtained from ground calibrations\cite{planck2011ther}.

Another source of 100mK plate temperature fluctuation is showers of particles resulting from interaction of very high energy particles with the payload. The rate of these event is about one per day. Showers of particles deposit energy in the 100mK plate but also directly in the detectors. these events are not discussed in this paper. For a detailed discussion see the LTD-15 Miniussi proceeding.

\subsection{Direct effect on detectors}

\begin{figure}
\begin{center}
\includegraphics[width=12cm, , keepaspectratio]{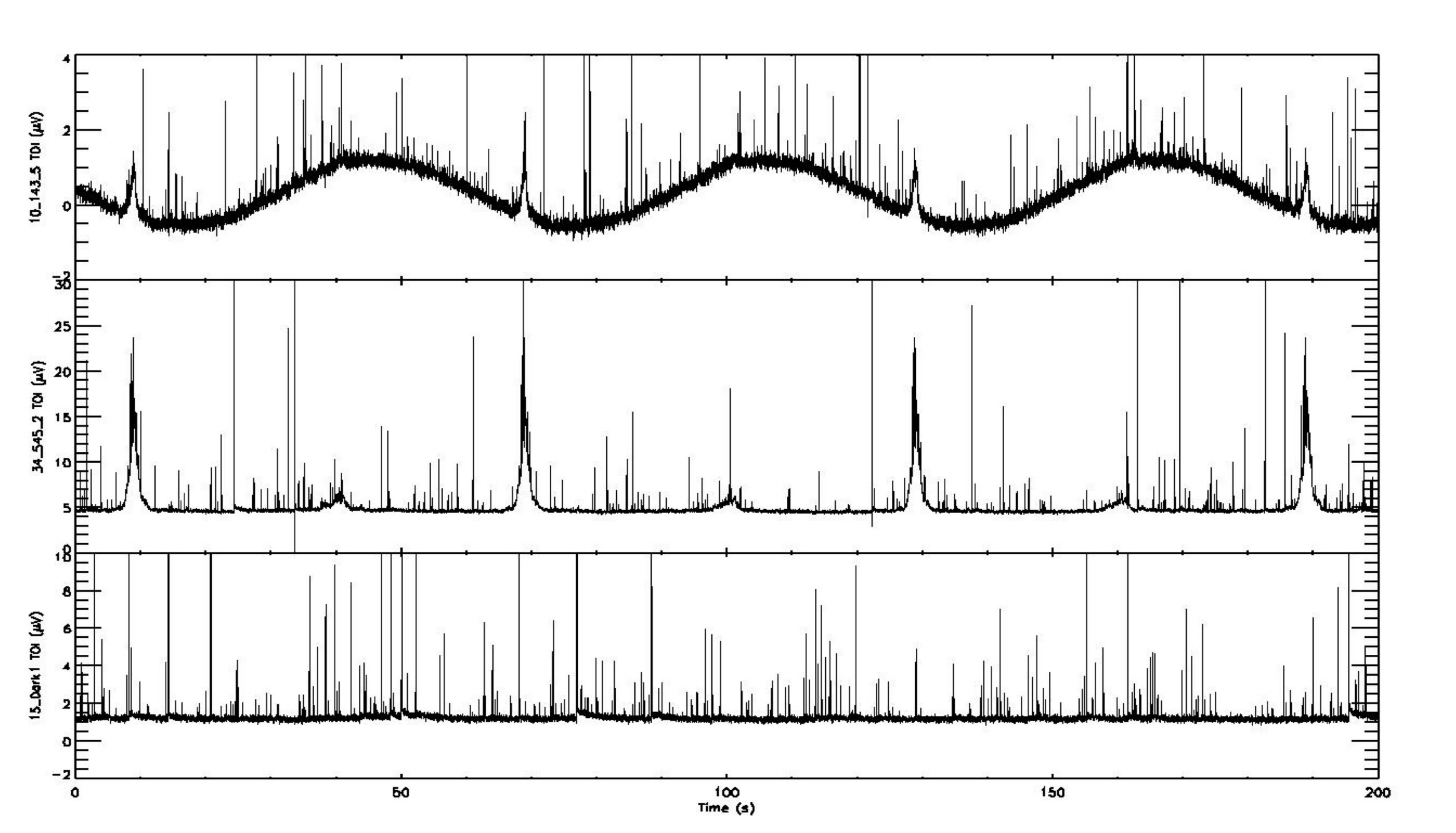}
\end{center}
\caption{Raw TOIs for three bolometers, 143GHz (top), 545GHz (middle), and a Dark1 bolometer (bottom) illustrating the typical behaviour of a detector at 143 GHz, 545 GHz, and a blind detector over the course of three rotations of the spacecraft at 1 rpm. At 143 GHz, one clearly sees the CMB dipole with a 60 s period. The 143 and 545GHz bolometers show vividly the two Galactic Plane crossings, also with 60 s periodicity. The dark
bolometer exhibits a nearly constant baseline together with a population of glitches from cosmic rays similar to those seen in the two upper panels. The typical maximum amplitude of a spike is between 100 and 500 mV depending on the bolometer.}
\label{fig:raw}
\end{figure}

\begin{figure}[t!]
\begin{center}
\includegraphics[width=10cm, , keepaspectratio]{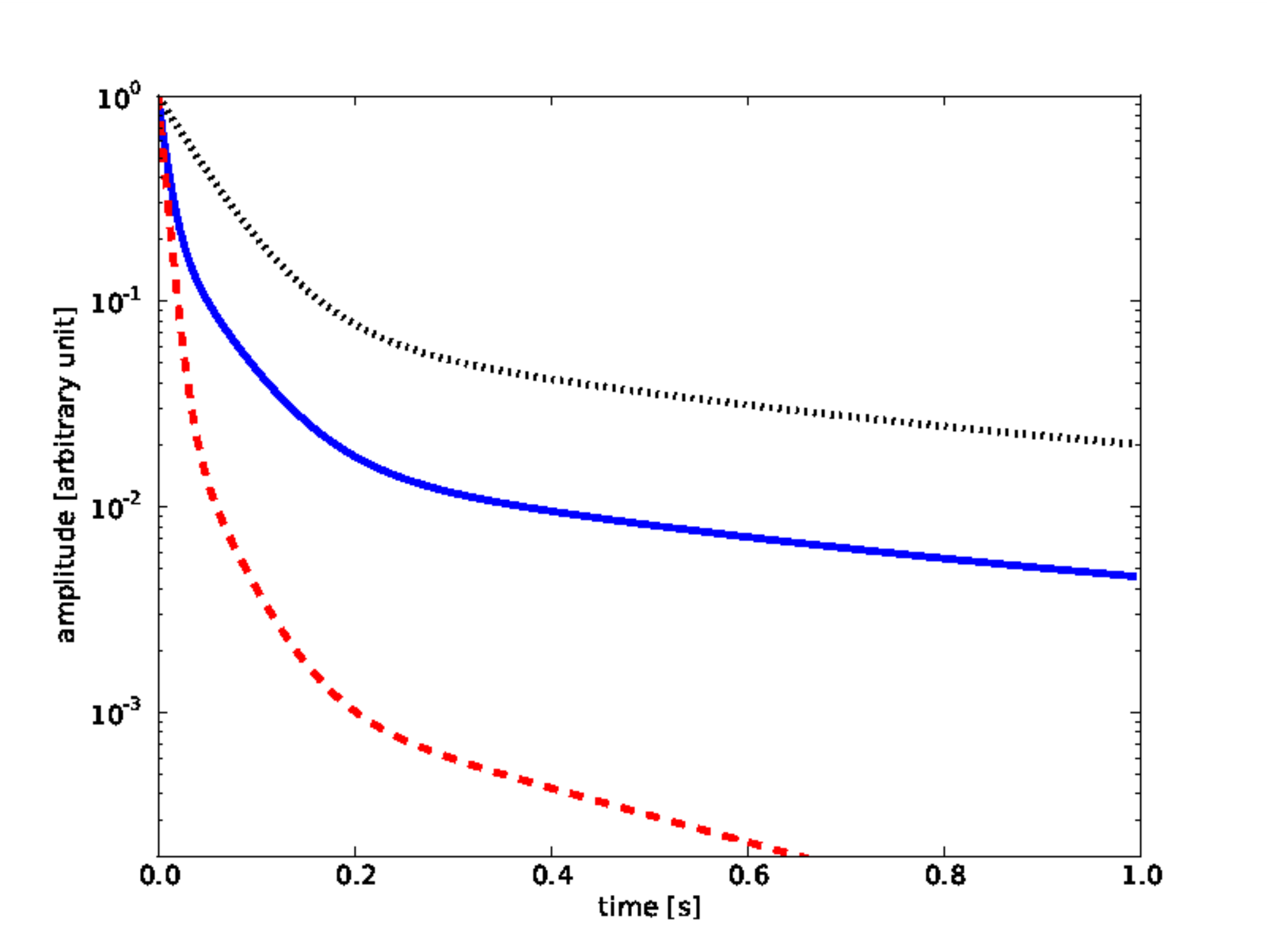}
\includegraphics[width=10cm, , keepaspectratio]{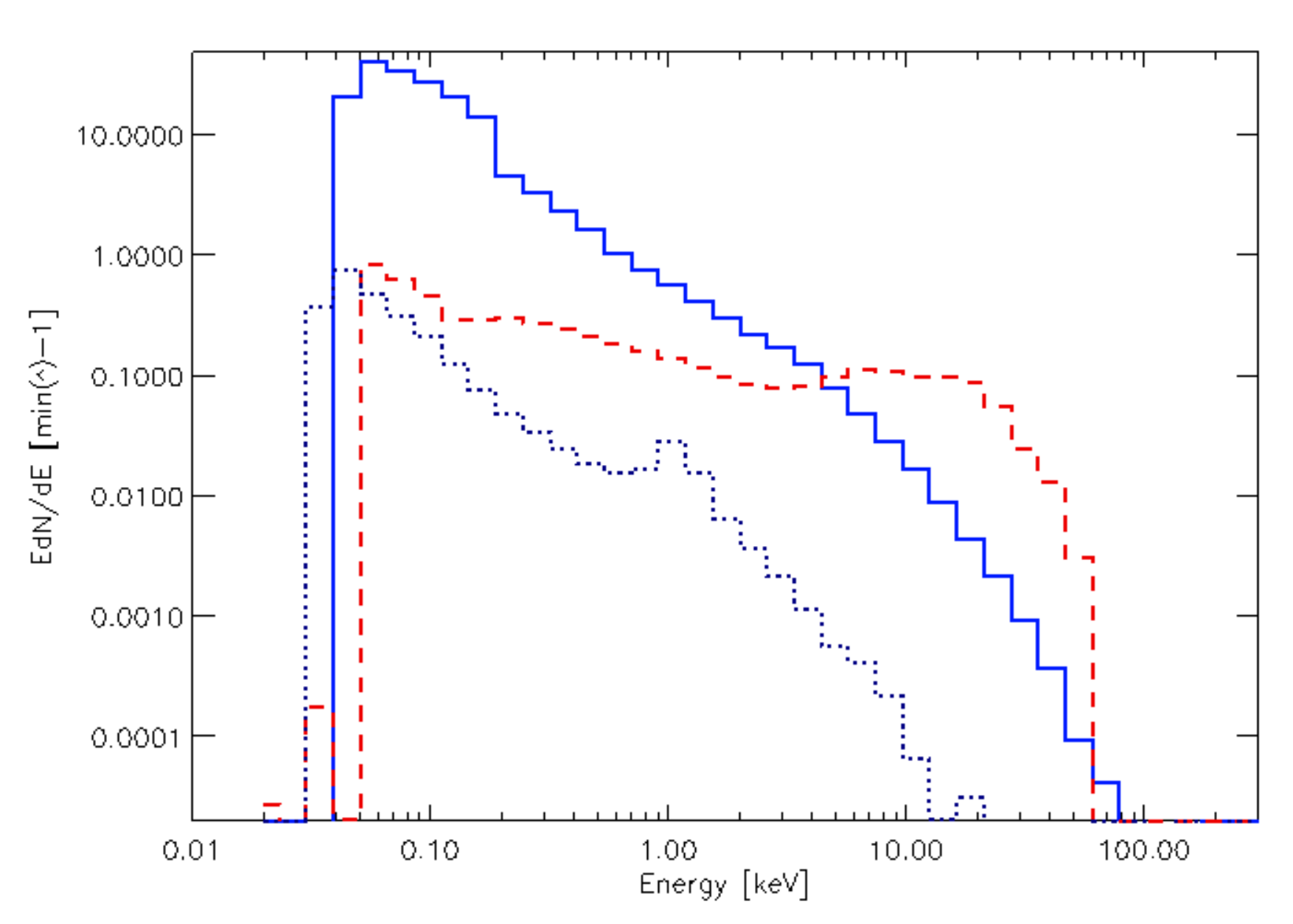}
\end{center}
\caption{Left: average short glitch template (red dashed curve), long glitch template (blue solid curve) and slow glitch template (dotted black curve) in the case of one PSB in- flight bolometer. Right: glitch energy distribution of the glitches observed by HFI in-flight for short (dashed red), long (solid blue) and slow (dotted black) in the case of one PSB in-flight bolometer; the x-axis represents the absorbed energy in the NTD.}
\label{fig:glitch}
\end{figure}

Glitches in raw data result from the direct impact of cosmic rays in the bolometer module. In Fig \ref{fig:raw} we present the raw data for three HFI detectors showing that the rate is quite conspicious (about 2 per second). 

The collaboration has identified three types of glitches with different shapes. 
The fitted templates and the spectra of these short, long and slow types of glitch are drawn in Fig \ref{fig:glitch}.

Short glitches present a fast decay (between 4 -- 10\,ms depending on
the bolometer) and a tail with an amplitude of fews percentage points relative to the largest amplitude.  
The tail shows intermediate time constants of the order of tens
of ms and a third time constant of about 1\,s with a relative amplitude
of 0.1\,\%.
Long glitches present the same fast decay as the short ones but a
large (about 10\,\%) relative amplitude tail showing intermediate to 
long time constants of tens of ms and a third time constant of about
1\,s with a relative amplitude of 0.1\,\%. 
Slow glitches show the same slow time constants but not the fast decay; they are present only in the foward PSB (in Planck focal plane PSBs are arranged in pairs sensitive to orthogonally-oriented
polarization).

– Slow and long glitch spectra exhibit similar shapes, so they
should share some parts of the of the energy deposition process.
– Long glitch spectra have a plateau below 1 keV followed by
a power-like spectrum and a smooth cut from about few hundreds of keV to 1 MeV.
– Short glitch spectra have a double structure with a power-law
below 10 keV followed by a bump very similar for
all bolometers (except 857 GHz ones) and then few events
above 100 keV (around one per day). Very high energy
events are rare but we have enough statistics to see such unprobable
signals due to cosmic ray crossing the grid along a
line.

\section{The origin of the excess HFI glitches : ground tests}\label{ground}

For an improved understanding of the origins of the glitches \cite{caserta} seen in
HFI flight data, we performed several ground-based tests on HFI spare
bolometers between the late 2010 and spring 2013. The aim
of the tests is to have a more complete view of the physical explanation of the
different families of glitches. These tests have been performed in
different configurations using 23\,MeV protons from a TANDEM
accelerator\footnote{http://ipnweb.in2p3.fr/tandem-alto/} and two radioactive $\alpha$ particle sources ($^{241}$Am, 
$^{244}$Cm and X-rays $^{55}$Fe to calibrate the signal). The list of all the performed tests with the main
characteristics of the setup is presented in Table \ref{tab1}.

\begin{table*} 
  \begin{center}
  \scalebox{0.9}{%
    \begin{tabular}{cccc}
      \hline
      & TANDEM acc. Tests &  $\alpha$ Test 1 & $\alpha$ Test 2  \\
      \hline  \hline	
      Place		 & IPN - Un. of Paris-sud & Neel Institute - Grenoble & IAS - Un. of Paris-sud \\
      
      Period &  Dec. 2010 & Nov. 2011--Apr. 2013 & June--Nov. 2011 \\
  
  Source	 &TANDEM accelerator  & $^{55}Fe$ isotope source   & $^{241}Am$ Source  \\
	 & $p^+ $ at 23\,MeV 	&  (X-rays 5.4\,keV, 1.3 kBq  & ($\alpha$ particles at 5.4\,MeV,   \\
         & 					&   produced in 2006) 	    &  3\,Bq)   \\
        & 					& $^{244}Cm$ Source  	    &  \\
        & 					& ($\alpha$ particles at 5.9\,MeV, about 1kBq) & \\
  Cryostat	 & N\'eel 100\,mK dilution & N\'eel 100\,mK dilution& IAS 100mK dilution  \\  
  
  Detectors	 &3 SWB 	& 1 SWB - 1 PSB& 1 SWB - 1 PSB  \\
  
  Read-Out El.& AC biased (Same as HFI) & AC biased (Same as HFI) & DC (dig. at 5\,kHz) \\
  \hline  \hline
    \end{tabular}
 }
  \end{center}
  \caption{List of all the performed tests with the principal characteristics of the setup.} \label{tab1}
\end{table*}

 
\section{Interpretation} 
\subsection{Short glitches origin}

\begin{figure}[b!]
\begin{center}
\includegraphics[width=10cm, , keepaspectratio]{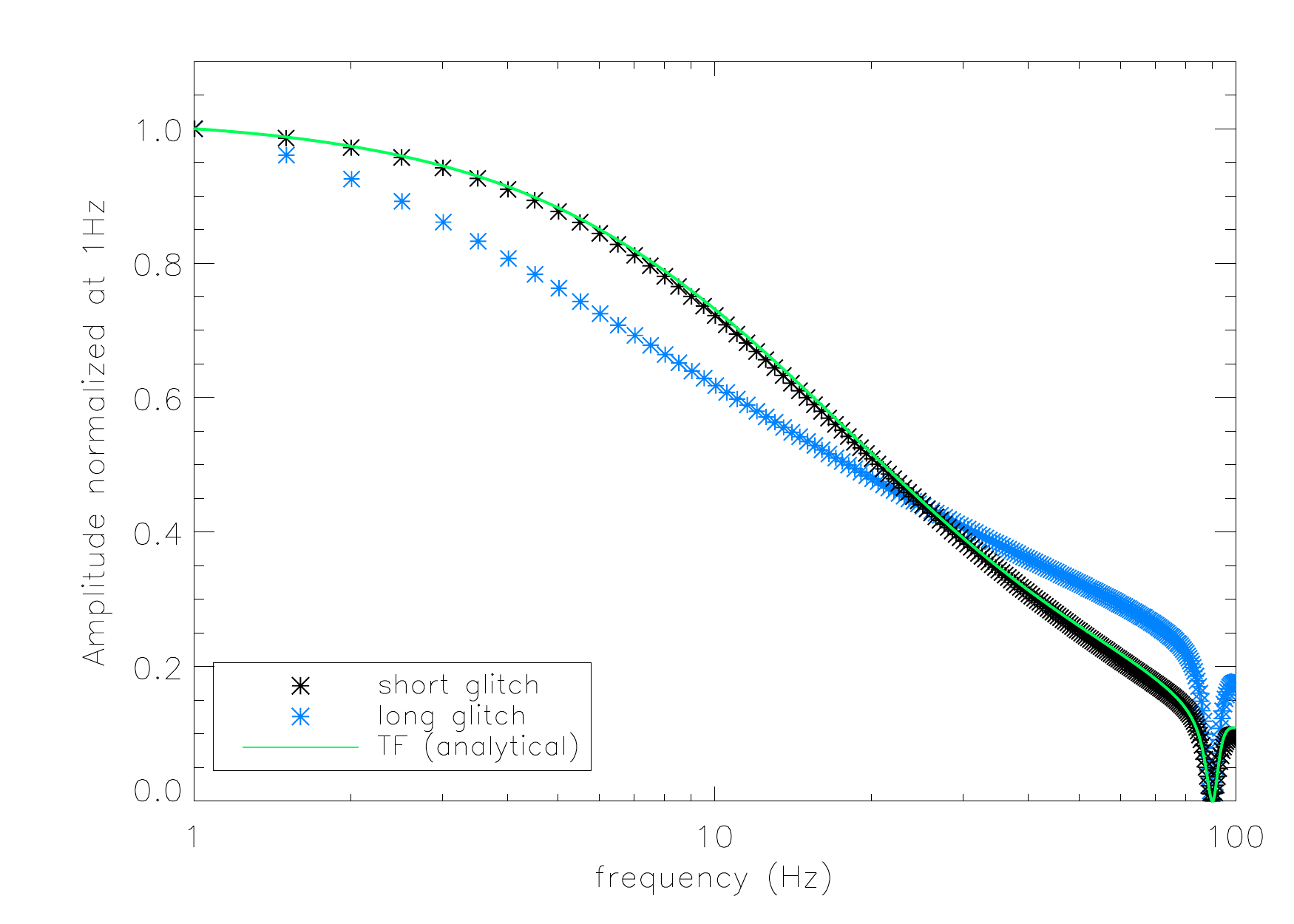}
\includegraphics[width=10cm, keepaspectratio]{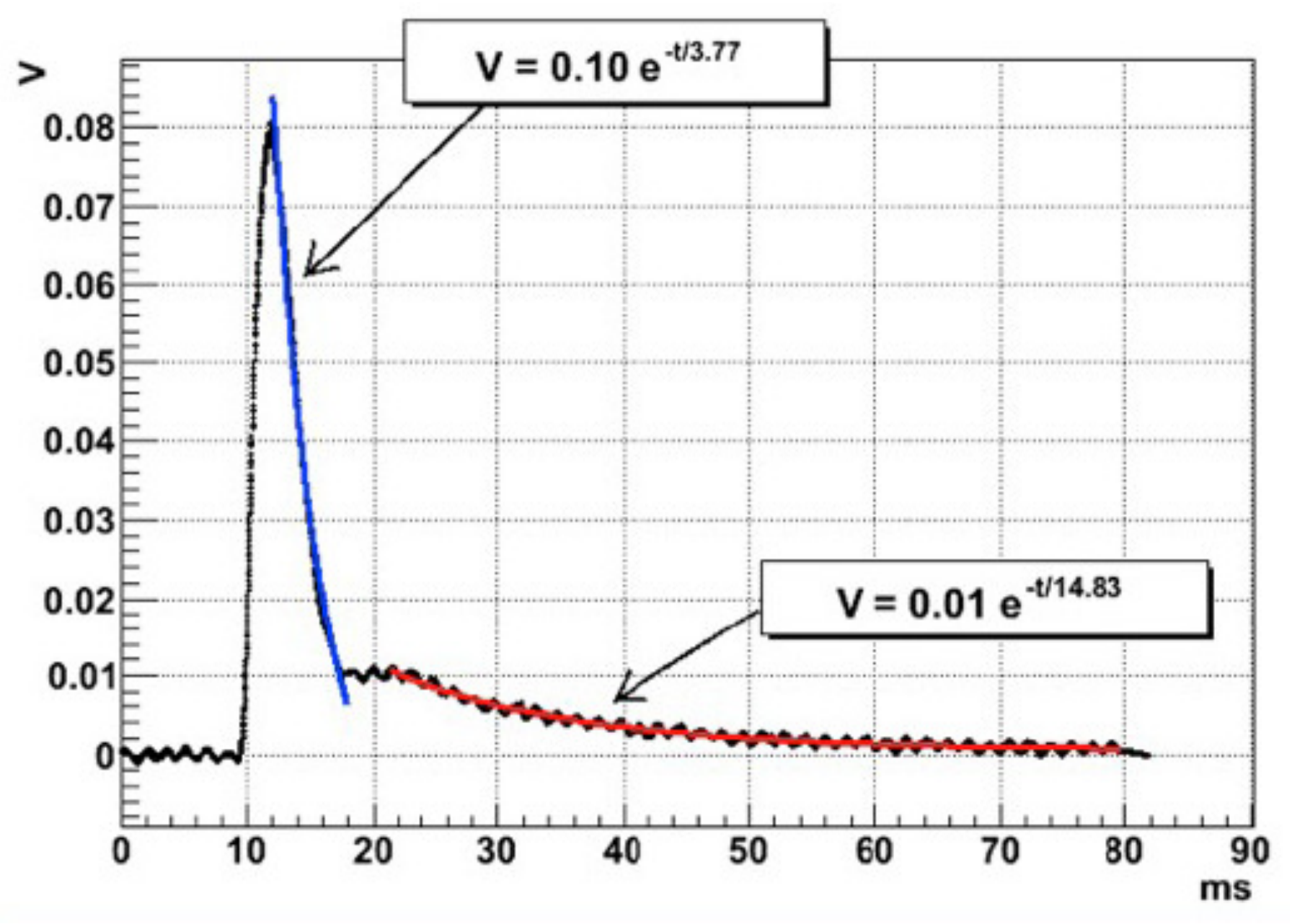}
\end{center}
\caption{Left: Comparison of the HFI Optical transfer function\cite{Planck2011perf} (solid green), short glitch templates (black stars) and long glitch template (blues stars). Right: glitch template built by stacking 466 events coming from impacts in the silicon wafer.}
\label{fig:ind_eff1}
\end{figure}

There is clear evidence that the short events are resulting from cosmic rays hitting the grid or the thermistor. Indeed, those events have a fast rising time and have a fast decay, and the transfer function (Fig \ref{fig:ind_eff1} left) built from the short glitch template is in good agreement with the HFI optical transfer function, so the energy must be deposited in the environment close to the thermistor. 
The energy distribution of the short glitches is very stable between bolometers and the data show a double structure\cite{Planck2013glitch}. The bump of the short glitch distribution, centred at about 20 keV, is completely consistent with the interaction between CRs and the NTD thermometer. In the low energy regime, the distribution corresponds to the interaction between the absorber and the CRs.
For PSB bolometers, the level of PSB-a/b glitch coincidence measured during ground-based testing is in good agreement with in-flight HFI PSB pair data, but in both cases it is greater than what we would expect from direct interactions only. We can suppose, therefore, that if a particle interacts with one grid, some delta electrons can be ejected, causing an increase to the rate of glitch coincidence between PSB pairs.

\subsection{Long glitches origin}

Two hypotheses have been put forward in order to explain the origin of the long glitches: the first was that the NTD thermometer is sensitive to a change in temperature of other
(larger) elements. The second was that a large part of the
glitches come from indirect interaction between CRs and bolometers;
for example, protons which interact very close to the surfaces of
materials surrounding the NTD+Grid (in particular copper) produce
electron showers able to propagate to the absorber+NTD.

We identify the long glitches as produced by cosmic rays hitting the silicon die. This was first indicated by the ground tests (Catalano et al. in prepration), showing that the NTD thermometer is sensitive to a temperature change of the silicon die (Fig \ref{fig:ind_eff1} right). The HFI ground-based calibration show a rate of events compatible with the cosmic rays flux at sea level over the silicon die surface and also that almost all these events are in coincidence between PSB-a and PSB-b. The understanding is the following: phonons generated by the event impact in the silicon die produce fast rising time of the Germanium temperature, which decays with the bolometer time constant. The slow part is the thermal response of the entire silicon die temperature rising and then falling as the heat conducts out from the die to the heat sink.

In order to reinforce this hypothesis, we have developed a toy model \cite{Planck2013glitch} considering the impact of cosmic rays at
the second Lagrange point with the silicon die. We start with a solid
square box made of silicon with the same equivalent surface of the
real silicon die. We consider that the side
of the square is much greater then the thickness of the silicon die.
The input of the model are:
\begin{itemize}
\item geometrical parameters of the bolometers;
\item stopping power function and density of the silicon die;
\item energy distribution of CRs at L2 \cite{pamela};
\end{itemize}

\begin{figure}[t!]
\begin{center}
\includegraphics[width=12cm, , keepaspectratio]{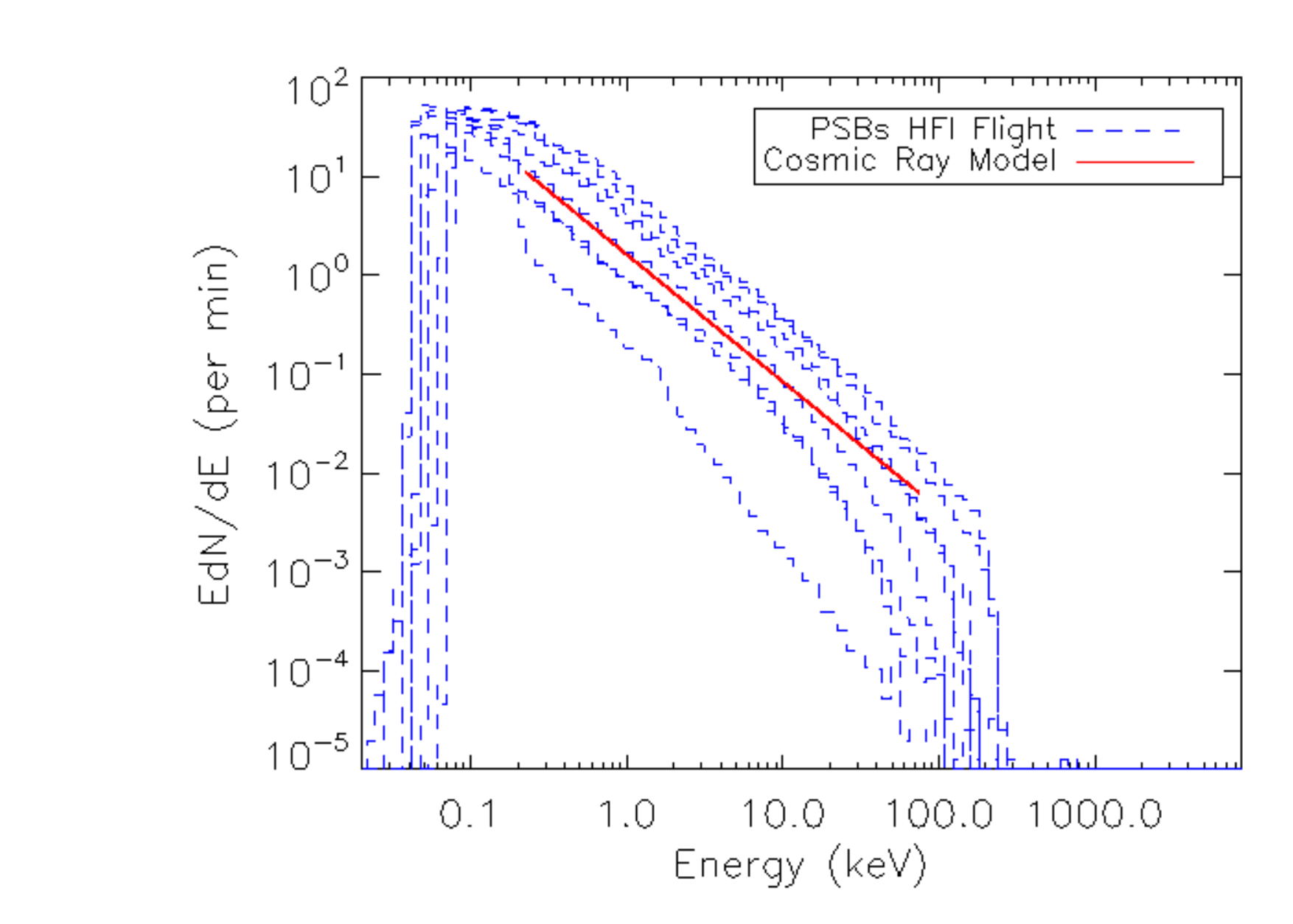}
\end{center}
\caption{Energy spectrum of some typical HFI in-flight bolometers (blue dashed lines) together with the predictions (red line).}
\label{fig:model}
\end{figure}

By integrating over the solid angle, the surface and the integration
time, we obtain an analytical equation for the number of events per unit of 
absorbed energy as:
\begin{equation}\label{cr_model}
  \frac{\Delta N}{\Delta E_{abs}} = \frac{4 \pi N_0 \Delta t E_{p^+o}}{(2\gamma+\beta-1) \cdot E_{O}^{\frac{\beta-1}{\gamma}}} \cdot E_{die}^{- \frac{\gamma + 1-\beta}{\gamma}},
\end{equation}
where $N_0$ is the amplitude of the spectrum of incoming protons or
alpha particles at L2, $\Delta t$ the integration time, $E_{p^+o}$ the
reference proton energy, $\gamma$ and $\beta$ are the power-law
indexes of the fit between the stopping power function and the energy
distribution of the proton at L2 respectively, $E_{O} = \rho_{sil}
\cdot d \cdot SP(E_{p^+o})$ is the reference absorbed energy for an
orthogonal impact, $\rho_{sil}$ the density of the silicon, $d$ the
thickness of the silicon die, $SP(E_{p^+o})$ the stopping power
function calculated at a reference proton energy, and $E_{die}$ is the
energy absorbed by the silicon die.

The predictions of this model for protons are
shown in Figure~\ref{fig:model}, together with the energy spectrum
measured in bolometer data. The power-law index fits well with the
flight data and the model is able to cover almost all the range of
energies. We conclude that in terms of rate and energy distribution
the thermal coupling between the silicon die and the NTD thermometer
likely explains the long glitches seen in HFI in-flight data.
The model shows the cut off of the long glitches distribution of the low signal is close to the noise level which is a critical point in terms of cosmological HFI goals.

\subsection{Slow glitches origin}

Slow glitches are the rarest events we see in terms of individual
glitches on the HFI in-flight bolometers. They affect only the
polarized PSBa bolometers with a rate of a few per hour in flight. The energy
distribution of the slow glitches shows the same power-law index of
the corresponding long glitches energy distribution (see Fig.\ref{fig:glitch}). 
In addition, the slow glitches share the template of the long glitches, 
but without the shortest time constant. These slow glitches were not reproduced during any of 
the HFI ground-based tests, both pre-launch with the HFI focal plane unit (FPU), and post-launch with flight-spare hardware.
In light of this fact, we are limited to putting forward
an hypothesis on the slow glitches origin that is consistent with our current results, but is without experimental confirmation.
The presence of the feed-through connecting the PSBa bolometer to its corresponding silicon die is 
the only difference between the PSBa bolometers and the other types of bolometers in HFI, i.e., both PSBb and SWB bolometers.  
The PSBa feed-through elements have a strong thermal coupling with the system
silicon die and gold pad. A proton hitting a PSBa feed-through, therefore, can heat
the corresponding silicon die to produce a heat diffusion from the silicon
die to the NTD thermometer.  This heating would be without the corresponding ballistic 
heat conduction associated with a silicon die/CR glitch event, and
therefore no fast time constant would be observed. The differences in the effective surface area of the
feed-throughs with respect to the corresponding silicon dies, of a factor of about 100, 
may explain the differences in the rate between the long glitches and
the slow glitches.


\section{Conclusion}

Understanding the origin and the features of glitches is of primary importance to manage systematic errors associated to the cosmic rays\cite{masi}. In addition to direct impact between the cosmic particles and the sensitive parts of the bolometers, it appeared that hits on the silicon die are also detectable and are a few tens time more numerous than the expected component. Moreover, the solar activity was extremely low at the beginning of the mission, and so the flux of cosmic ray was unusually high.

The effort done by the collaboration allowed to obtain data of excellent quality to make the maps but 20 to 12\% of the samples are discarded due to glitch contamination\cite{planck2013p01}.
To minimise the cosmic flux, for a balloon or a space experiment using low temperature devices, it is preferable to try to avoid deep solar minimum, in particular the lowest one which occurs with a period of 100 years. But then, the experiment can suffer from solar flares so a trade-off has to be done. The crucial point, which is more manageable than the solar meteorology, would be the improvement of the isolation between the die and the sensitive part of the bolometer by an order of magnitude. Then the rate of glitches coming from the bolometer and the silicon die will be comparable, even during a period of small solar activity.
The implications of this work for future satellite come from the need to improve by at least one order of magnitude the Noise Equivalent Power of new space experiments (from $10^{-17}$ $W$ / $\sqrt{Hz}$ to $10^{-19}$ $W$ / $\sqrt{Hz}$). This can be reached
by increasing the focal plane coverage, using thousands of Background Limited Instrument Performance (BLIP) contiguous pixels. Each pixel of these arrays must be micro-machined starting from a common substrate. For these raisons, the influence of cosmic rays for future space detectors, must be taken into account from the very first phases of the design in parallel with all the other characteristics like NEP and time response. In particular, beam testing should be planned to study irradiation on whole array of detectors (pixels, substrate and housing).



\end{document}